\documentclass[%
 reprint,
superscriptaddress,
groupedaddress,
 amsmath,amssymb,
 aip, jcp,
floatfix,
]{revtex4-1}

\usepackage{graphicx}%
\usepackage{dcolumn}%
\usepackage{bm}%
\usepackage{graphicx}%
\usepackage{dcolumn}%
\usepackage{bm}%
\usepackage{xcolor}
\usepackage{mhchem}
\usepackage{gensymb}
\usepackage{ulem}
\usepackage{comment}
\newcommand{\si}{Appendix}

\newcommand{\gt}[1]{{\color{black} #1 }}
\newcommand{\avg}[1]{\ensuremath{\left<#1\right>}}

\newcommand{\pd}[2]{\ensuremath{\frac{\partial {#1}}{\partial {#2}}}}
\newcommand{\pds}[2]{\ensuremath{\frac{\partial^2 {#1}}{\partial {#2}}}}
\newcommand{\bigo}[1]{\ensuremath{\mathcal{O}\left({#1}\right)}}

\newcommand{\xyz}{\ensuremath{(x,y,z)}}
\newcommand{\tht}{\ensuremath{\theta}}
\newcommand{\ph}{\ensuremath{\phi}}
\newcommand{\tp}{\ensuremath{(\theta,\phi)}}
\newcommand{\htp}{\ensuremath{(\hat{\tht},\hat{\ph})}}
\newcommand{\htht}{\ensuremath{\hat{\tht}}}
\newcommand{\hph}{\ensuremath{\hat{\ph}}}

\newcommand{\drtheta}{\ensuremath{\partial r / \partial \tht}}
\newcommand{\drphi}{\ensuremath{\partial r / \partial \ph}}
\newcommand{\tpsi}{\ensuremath{\widetilde{\psi}}}
\newcommand{\tpsis}{\ensuremath{\widetilde{\psi}_{\text{s}}}}
\newcommand{\tpsil}{\ensuremath{\widetilde{\psi}_{\text{l}}}}
\newcommand{\tpsio}{\ensuremath{\widetilde{\psi}_{\text{0}}}}
\newcommand{\rhos}{\ensuremath{\rho_\text{s}}}
\newcommand{\rhol}{\ensuremath{\rho_\text{l}}}
\newcommand{\rhoo}{\ensuremath{\rho_\text{0}}}
\newcommand{\mr}{\ensuremath{\bar{r}}}

\begin{document}

\preprint{APS/123-QED}

\title{Classical nucleation theory predicts the shape of the nucleus \\ in homogeneous solidification}

\author{Bingqing Cheng}
\email{bc509@cam.ac.uk} 
\affiliation{
TCM Group, Cavendish Laboratory, University of Cambridge, J. J. Thomson Avenue, Cambridge CB3 0HE, United Kingdom
and \\
Trinity College, University of Cambridge, Cambridge CB2 1TQ, United Kingdom}%

\author{Michele Ceriotti}
\affiliation{Laboratory of Computational Science and Modeling, Institute of Materials, {\'E}cole Polytechnique F{\'e}d{\'e}rale de Lausanne, 1015 Lausanne, Switzerland}%

\author{Gareth A. Tribello}
\affiliation{Atomistic Simulation Centre, School of Mathematics and Physics, Queen's 
University Belfast, Belfast, BT7 1NN
}%

\date{\today}%

\begin{abstract}
Macroscopic models of nucleation provide powerful tools for understanding activated phase transition processes. These models do not provide atomistic insights and can thus sometimes lack material-specific descriptions. Here we provide a comprehensive framework for constructing a continuum picture from an atomistic simulation of homogeneous nucleation. We use this framework to determine the equilibrium shape of the solid nucleus that forms inside bulk liquid for a Lennard-Jones potential. From this shape, we then extract the anisotropy of the solid-liquid interfacial free energy, by performing a reverse Wulff construction in the space of spherical harmonic expansions. We find that the shape of the nucleus is nearly spherical and that its anisotropy can be perfectly described using classical models.
\end{abstract}

\pacs{Valid PACS appear here}%
\maketitle

\section{\label{sec:level1}Introduction}

Nucleation is an essential component of many technological and natural processes~\cite{sosso2016crystal}.  A better understanding of nucleation would help us to understand precipitation in the atmosphere, the casting of metals~\cite{flemings1974solidification}, the formation of amyloid plaques in the brain~\cite{vsaric2014crucial},
the formation of biominerals and how to preserve bodily fluids such as blood and spinal fluid so that they can be used in transfusions~\cite{veis2013biomineralization}.  
Performing experiments to determine what occurs during a nucleation event is fraught with difficulties~\cite{pouget2009initial,zhou2019observing}, however, because of the small length and timescales over which nucleation takes place. 
In particular, it has only recently become possible to determine
the three-dimensional atomic structure and the dynamics of small nuclei at the early stage of nucleation~\cite{zhou2019observing}.
Consequently, a great deal of theoretical and simulation work has been performed to understand how crystals nucleate and grow~\cite{sosso2016crystal}.  

One modelling technique that can be used to understand nucleation is molecular dynamics (MD).  This technique models the interactions between each of the atoms or molecules in the nucleus and the surrounding melt/solution~\cite{sosso2016crystal} explicitly.  
This approach thus takes account of material-specific information
and generates atomistic insight. When this method is applied, however, a vast amount of information including all atomic coordinates is typically generated, and sophisticated data analysis methods are therefore required to interpret the results~\cite{espinosa2016seeding,cheng2017gibbs,prestipino2018barrier}.

Classical nucleation theory (CNT) and phase-field models~\cite{cahn59jcp} are another set of tools that can be used to understand and model the process of nucleation.  
Unlike MD, these classical models do not include an explicit treatment of the individual particles that constitute the crystal and its surroundings.  
In CNT the total volume is instead partitioned into a nucleus with bulk properties and the surrounding metastable bulk phase.  
Meanwhile, in the simplest phase-field models, the system is described using a mean-field order parameter that is given a value at each point in the volume~\cite{cahn59jcp, granasy2002interfacial}.
Generally speaking, the total free energy of the system in these classical models consists of a sum of volume terms due to the bulk phases and surface terms that are due to the interfaces between the various phases.
The results from these classical models are thus easy to interpret as they provide one with a solid physical understanding.  Furthermore, these models can be coupled with macroscopic models of heat and mass transport to predict the outcomes of solidification in a specific technological application.

Given that these two theoretical approaches for studying nucleation are complementary, it should come as no surprise to find that numerous attempts have been made to extract the parameters for the classical models from molecular dynamics simulations. For example,
Ref.~\citenum{cheng2015solid} and Ref.~\citenum{cheng2017gibbs} describe a thermodynamic framework that uses a Gibbs dividing surface construction to extract interfacial free energies for planar and curved solid-liquid interfaces from MD simulations that can be used in expressions based on classical nucleation theory.  In using these methods, however, one must assume that the nucleus has a spherical shape or -- alternatively -- that the interfacial free energy $\gamma$ for planar interfaces oriented along high-symmetry directions provide sufficient information to approximate the anisotropy of $\gamma$ and, by extension, to estimate the shape of the nucleus through a Wulff construction. Another reason for linking  atomistic and continuum models is thus to  
test the limitations of the macroscopic descriptions of nucleation that appear in phase-field and continuum models.
In this paper, we, therefore, extend the construction in Refs.~\citenum{cheng2015solid,cheng2017gibbs} and extract the average shape of the nucleus from a molecular dynamics simulation directly.  \gt{In Ref. \citenum{cheng2017gibbs} we extracted this shape by performing a Wulff construction using values for the surface tensions that were extracted from simulations of planar interfaces.}  We thus finish this work by performing a quantitative comparison between the shape we extract \gt{from our simulations of a three-dimensional nucleus and the shape that was predicted using the Wulff construction.  We find a level of quantitative agreement between the shapes extracted using these two methods that indicates that it is appropriate to use  classical nucleation to describe this particular system.  We furthermore argue that similar comparisons of the shapes extracted using the methods that we have introduced in this work and our previous works \cite{cheng2015solid,cheng2017gibbs} can now be used to check if the nucleation mechanisms observed in other physical systems follow the classical theory.  } 

The details of the methodology are explained in section \ref{sec:method}.  Section \ref{sec:details} then provides details on the system of Lennard-Jonesium that we have simulated.  Section \ref{sec:analysis} describes several subtle issues in the analysis in more detail and finally, in section \ref{sec:results}, the approach is used to determine the shape of a nucleus of Lennard-Jonesium.  As discussed in the previous paragraph, this analysis demonstrates that, for this system, the shape of the nucleus that is extracted from a molecular dynamics simulation is consistent with a prediction \gt{of the shape that is obtained using a Wulff construction parameterized using values of the surface tension extracted from simulations of planar interfaces. In other words, the behavior of the nucleus in our molecular dynamics simulations is entirely consistent with the predictions of the classical models.}
         
\section{\label{sec:method} Theory}

\subsection{Phase-field representations}

In this section, the trajectory from a molecular dynamics simulation of a system that contains one or more solid nuclei inside a bulk liquid phase is considered. 
The energetic barrier associated with the formation of an interface between the two phases ensures that large crystalline nuclei rarely form.
An external bias potential, $U_{\textrm{bias}}(\Psi)$ that is a function of an extensive thermodynamic variable, $\Psi$, can be used to force a nucleus of any size to form or grow in a biased simulation, however.  Furthermore, because $\Psi$ is extensive, it can be decomposed into a sum of individual atomic contributions, i.e. $\Psi=\sum_i\psi_i$ where the sum over $i$ here runs over all the atoms in the system. 
These atomic contributions are useful because they allow the atomistic representation of the system to be converted to a phase field picture~\cite{bald+17jpcm} using:
\begin{equation}
\tpsi\xyz = \sum_i \psi_i K\left(x - x_i, y - y_i, z - z_i \right).
\label{eqn:pfield}
\end{equation}
In this expression $K$ is a three-dimensional, normalized kernel function such as a Gaussian, and $(x_i,y_i,z_i)$ is the position of atom $i$.  Furthermore, we can perform the above conversion for a single trajectory frame or, as long as we ensure that the atomic coordinates are aligned to some common reference frame, we can compute the ensemble average $\avg{\tpsi\xyz}$ over multiple trajectory frames.

\begin{figure}
    \centering
    \includegraphics[width=0.4\textwidth]{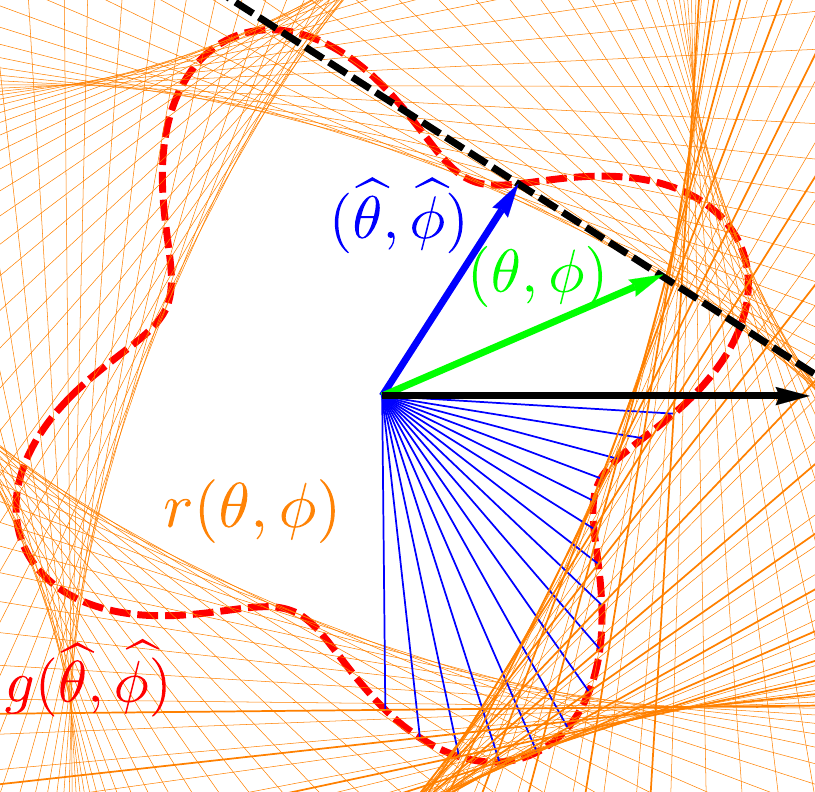}
    \caption{Schematic illustrating a Wulff construction. $r \tp$ (the area cut out by the orange tangent lines) indicates the equilibrium shape of the nucleus.
    $g\htp$ (dashed red curve), meanwhile, is the surface energy per unit area for the surface whose normal points along $\htp$.  Two different sets of polar coordinates are used in these expressions because the nucleus is not spherical, so the vectors normal to the surface do not point along these radial vectors.
    As such, the surface point $r \tp$ has surface energy per unit area $g\htp$.}
    \label{fig:Wulfff}
\end{figure}

We use $\tpsis$ and $\tpsil$ to indicate the average values for the phase-field density inside the bulk solid and liquid phases, respectively.
For a solid-liquid system, 
$\tpsi\xyz$ takes values that are closer to $\tpsis$ in the parts of the simulation box, where many atoms have a solid-like structure.  By contrast, in parts of the box where the structure is liquid-like $\tpsi\xyz$ takes values that are closer to $\tpsil$.  
One way to determine the location of the surface that separates the two phases is to place a constant threshold value on the value of the atomic order parameter and to thereby distinguish the solid-like atoms from the liquid-like atoms. This approach is commonly used in tandem with capillary fluctuation methods~\cite{hoyt2001method}. In this work, however, we want to be able to locate the surface even when the atomic order parameters exhibit large fluctuations. We, therefore, use a method that is analogous to the Willard--Chandler construction \cite{chandler-1,chandler-2,bald+17jpcm}.  In particular, the location of the interface between the solid and liquid phases is found by searching for a manifold of points that satisfy:
\begin{equation}
\tpsi\xyz - \tpsio = 0
\label{eqn:contour}
\end{equation}
A different choice for $\tpsio$ here inevitably leads to different locations for the interface. For a planar interface at the coexistence temperature, this is unimportant as the precise location of the interface is irrelevant. By contrast, for a finite nucleus, any shift in the position of the interface changes the size of the nucleus, and thus changes the excess interface free energy.
One way to remove this arbitrariness is to enforce a zero-excess condition for the extensive quantity, $\Psi$.  
This condition amounts to finding the value of $\tpsio$ for which the following integral over the whole volume, $V$, of the system is equal to $\Psi$:
\begin{equation}
     \Psi = 
     \iiint_V \left[ \tpsis H(\tpsi\xyz-\tpsio)
     + \tpsil H(\tpsio-\tpsi\xyz)
     \right],
     \label{eqn:gibbs}
\end{equation}
In this expression
$H(\ldots)$ is a Heaviside function,
and we assume that $\tpsis > \tpsio > \tpsil$.  
In fact,
eqn.~\eqref{eqn:gibbs} can be seen as an extension of the Gibbs dividing surface based on the extensive quantity $\Psi$~\cite{cheng2015solid,cheng2016bridging}.  It is, however, not necessary to construct a phase-field representation based on the extensive quantity $\Psi$ that was used to bias the molecular dynamics simulations as other extensive quantities can be used in place of $\Psi$ in  Eqn.~\eqref{eqn:pfield}.
For instance, if one sets all the $\psi_i$ values in Eqn.~\eqref{eqn:pfield} to one the resulting phase field is just the number density of atoms $\rho\xyz$,
and the corresponding extensive quantity for the whole system is just the total number of atoms $N$.
In this case, $\tpsis$ and $\tpsil$ in Eqn.~\eqref{eqn:contour} are the number density of atoms in bulk solid and bulk liquid, $\rhos$ and $\rhol$, respectively.
Furthermore, as we will later explain in the results section, when applying Eqn.~\eqref{eqn:contour} to find the location of the interface the value of $\rhoo$ can be set so as to ensure that 
$N=\iiint_V \left[ \rhos H(\rho-\rhoo)+ \rhol H(\rhoo-\rho)\right]$.
This surface contour, like the Gibbs dividing surface, has zero surface excess for the volume and  is referred to as the equimolar dividing surface in the literature~\cite{tolman1948consideration,cheng2017gibbs,cheng2018communication}.  This surface is unique because one can use it to map the solid-liquid system into a solid part which has a density that is the same as the density of bulk solid $\rhos$, a liquid part which has a density that is the same as the density of bulk liquid $\rhol$,
and a surface term with zero surface excess for the volume.

\subsection{Nucleus shape, free energy, and Wulff construction}

For three dimensional nucleation, the manifold of points at which the condition in Eqn.~\eqref{eqn:contour} is satisfied has the same simple topology as a sphere. We can thus express the equilibrium surface of the nucleus, $r \tp$, as a function of the polar angles.  As Figure~\ref{fig:Wulfff} shows, the normal vector to the surface $r \tp$ does not, in the general case, have to point in the same direction as the radial vector $\tp$.  
The orientation of these normal vectors can still be specified in spherical polar coordinates, however, so we introduce the symbols $\htht$ and $\hph$ to describe the direction, $\htp$, in which the normal to the surface $r \tp$ points when the polar angles are equal to $\tht$ and $\ph$.  
With this nomenclature in place we can make the connection with classical nucleation theory by writing the free energy of a nucleus as:  
\begin{equation}
\begin{aligned}
f  = & \int_{-\pi}^\pi \int_0^\pi \left[ r^2 \tp g\htp\sqrt{1+c\tp} \right] \sin\tht \textrm{d}\tht \textrm{d}\ph \\ 
& - \frac{1}{3} \int_{-\pi}^\pi \int_0^\pi \mu r^3 \tp \sin\tht \textrm{d}\tht \textrm{d}\ph \\  
\textrm{where} & \quad c\tp = 
\frac{1}{r^2}\left[ \left(\pd{r}{\tht} \right)^2 +  \frac{1}{\sin^2(\tht) } \left(\frac{\partial  r}{\partial \ph} \right)^2 \right]
\end{aligned}
\label{eqn:f}
\end{equation}
This expression is composed of two terms: the first of which is a surface term and the second of which is a bulk term.  Consequently, $g\htp$ is the surface energy per unit area for the surface whose normal points along $\htp$.
$\mu$ is then the per-unit-volume chemical potential of the solid relative to the bulk liquid.  This quantity can also be regarded as a Lagrange multiplier that enforces a constraint on the volume of the nucleus.

Generally speaking, 
the volume-specific chemical potential $\mu$ is not independent of cluster size,
because, although the per-particle chemical potential difference between the two phases is constant,
the density of the cluster can change with its size.  To make $\mu$ independent of the size of the cluster, one must, therefore, ensure that the size and density of the cluster are independent.  By using the equimolar surface, that was introduced in the previous section, \gt{we ensure that the nucleus always has the same molar volume and density as the bulk solid.  We thus use this surface in all our analyses to ensure that the volume-specific chemical potential difference is independent of the cluster size.} 

The equilibrium surface for a fixed size nucleus must correspond to a minimum in the free energy at which the first-order derivatives of Eqn.\eqref{eqn:f} with respect to $r$, $\theta$ and $\phi$ vanish.
As shown in the \si{}, computing these derivatives and manipulating the resulting simultaneous equations leads to the following expression:
\begin{equation}
g\htp = \frac{\mu  r \tp }{2 \sqrt{1+c\tp} }
\label{eqn:Wulff}
\end{equation}
which relates the equilibrium surface of the nucleus, $r \tp$, and the surface energy per unit area, $g\htp$.  This expression is fully consistent with the result that one would obtain using the geometric arguments behind the reverse Wulff construction \cite{dobrushin1992wulff}.  Furthermore, for spherical nuclei $\dfrac{df}{d r}=0$ and 
$c\tp=0$ so Eqn.~\eqref{eqn:Wulff} reduces to $\gamma = {\mu  r }/{2}$ which is the familiar relationship from classical nucleation theory that relates the critical radius $r$ and the isotropic surface energy, $\gamma$.
\gt{
It is important to note that, although Eqn.~\eqref{eqn:Wulff} is equivalent to CNT when it is applied to critical nuclei, this equation can also be used to describe pre-critical and post-critical nuclei.
This equation can be used in all these cases because, within it, $\mu$ serves as a Lagrange multiplier that enforces a constraint on the volume of the nucleus. This constraint can be applied by placing an external bias on the Hamiltonian for the system using methods such as metadynamics~\cite{laio-parr02pnas} or umbrella sampling~\cite{torrie1977nonphysical}.  Alternatively, in undercooled conditions, the volume is constrained because the bulk solid and bulk liquid have different chemical potentials.  Under such conditions it thus the balance between the surface and bulk terms in equation \ref{eqn:f} that determines the size of the nucleus. 
Critically, however, undercooling and bias potentials that act on the size of the nucleus affect the size of the nucleus only.  These factors do not affect the shape.  Consequently, if one normalizes both sides of Eqn.~\eqref{eqn:Wulff} by
dividing both sides by a factor of $\gamma=\mu r / 2$ one ends up with a measure of the shape that is dimensionless and thus valid for all nucleus sizes.
Once it is made dimensionless, Eqn.~\eqref{eqn:Wulff} even applies to the case where the chemical potential difference $\mu$ diminishes.  In the remainder of this manuscript we therefore always work with the normalized and dimensionless shape $r/\mr$ and the normalized and dimensionless interfacial free energy $g/\gamma$.}

Eqn.~\eqref{eqn:Wulff} can be simplified by assuming the nucleus is close to spherical and that $c\tp$ is thus small enough for $\frac{1}{\sqrt{1+c\tp}}$ to be replaced by its first-order Taylor series expansion around $c=0$.  Making this substitution, replacing $c\tp$ by the definition of this quantity from Eqn.~\eqref{eqn:f}  following expression:
\begin{equation}
    \dfrac{g\htp}{\gamma} = 
\frac{1}{\mr} 
\left\{ r - \frac{1}{2 r}\left[ \left(\pd{r}{\tht} \right)^2 +  \frac{1}{\sin^2(\tht) } \left(\frac{\partial  r}{\partial \ph} \right)^2 \right] \right\}.
\label{eqn:inter-g}
\end{equation}
We 
have used the shorthand $r$ for $r \tp$
in this expression and we have 
divided the left-hand side by the isotropic surface energy, $\gamma$,
and the right-hand side by the average radius of the nucleus $\mr$ which can be computed using
$(4\pi/3){\mr}^3=\int_{-\pi}^\pi \int_0^\pi r^2  \sin\tht \textrm{d}\tht \textrm{d}\ph$ so as to make both sides of the equation dimensionless.

Eqn.~\eqref{eqn:inter-g} is difficult to use because 
inserting a point on the nucleus surface, $r \tp$ into it
gives the value of the surface energy $g\htp$ along $\htp$ rather than along $\tp$,
and because the factor of $r$ in the denominator makes the equation nonlinear.
As shown in the \si{}, however, by using the assumption that the nucleus is close to spherical once more we can replace the left-hand side of Eqn.~\eqref{eqn:inter-g} with its Taylor expansion around $\tp$.  
If we then also replace the factor of $r$ in the denominator by its Taylor expansion around $\mr$,
and perform a change of variables between $\tp$ and $\htp$, 
we arrive at the following linearized expression that connects the equilibrium shape of the cluster with the excess free energy for the surface whose normal points along $\tp$:
\begin{equation}
 \dfrac{g\tp}{\gamma} = 
\frac{r}{\mr} + \frac{1}{2}\left[ \left(\frac{\partial (r/\mr) }{\partial \tht} \right)^2 +  \frac{1}{\sin^2(\tht) } \left(\frac{\partial (r/\mr) }{\partial \ph} \right)^2 \right] 
\label{eqn:final-g}
\end{equation}

\subsection{Reverse Wulff construction in the Spherical Harmonic basis}
\label{sec:final-method}

Eqn.~\eqref{eqn:final-g} is a central result for the present paper. In what follows we will discuss how to effectively exploit it when analysing atomistic trajectories.  The first step in this analysis is to obtain the average shape $ r \tp$ of the \gt{single crystal nucleus that is present in our simulation cell} by calculating an ensemble average from a molecular dynamics simulation.  To do so we first find the $\{x_{\textrm{c}},y_{\textrm{c}},z_{\textrm{c}}\}$ coordinates for the center of the nucleus for each frame of our molecular dynamics trajectory.  \gt{These coordinates are found using the formula below so as to account for the periodic boundary conditions in our simulation setup:} 
\begin{equation}
x_{\textrm{c}} = \frac{L_x}{2\pi} \arctan\left[ \frac{\sum_i \Theta(\psi_i) \sin\left( \frac{2 \pi x_i}{L_x} \right) }{ \sum_i \Theta(\psi_i) \cos\left( \frac{2 \pi x_i}{L_x} \right)} \right]
\label{eqn:center}
\end{equation}
where $L_x$ is the length of the $x$ axis of the simulation box and $x_i$ is the $x$ coordinate of the $i$th atom.  $\Theta(\psi_i)$, meanwhile, is a switching function that acts on the value of the the order parameter for the $i$th atom.  $\Theta(\psi_i)$ is thus one when $\phi_i$ takes its solid value and is zero otherwise. We then center all the frames in our trajectory on this reference point.  \gt{There is no need to rotate each configuration to the same reference frame because, as discussed in section \ref{sec:details}, the bias is a function of a rotationally variant order parameter.  Consequently, the nucleus that forms in our simulations always has one particular orientation in the lab frame.} 

Once all the frames are aligned we then either compute $\tpsi\xyz$ for each trajectory frame or the ensemble average for $\tpsi\xyz$ which is defined in Eqn.~\eqref{eqn:pfield}. Once $\tpsi\xyz$ has been computed we can then search radially outwards from the origin, which is at the position that was determined using Eqn.~\eqref{eqn:center}, for a set of points that satisfy Eqn.~\eqref{eqn:contour}.  In particular, and in order to have a set of grid points that are approximately evenly spaced over the surface of a sphere, we generate a set of $M$ vectors using the Fibonacci sphere algorithm \cite{fibonacci-sphere}.  The radial vectors $r^{(i)}$ that we search for the location of the contour on along are thus:
$$
\begin{aligned}
x^{(i)}   = \frac{2 i + 1 - M}{M} & \quad
y^{(i)} = r^{(i)}\sin(\ph^{(i)}) \quad z^{(i)} = r^{(i)}\cos(\ph^{(i)}) \\
\textrm{where} \quad (r^{(i)})^2 & = 1 - (x^{(i)})^2 \quad 
\textrm{and} \quad \frac{1+\sqrt{5}}{2} \ph^{(i)}  =  2\pi i
\end{aligned}
$$
where $M$ is a number that is in the Fibonacci sequence and where $\frac{1+\sqrt{5}}{2}$ is the Golden Ratio.  If $\tpsi\xyz$ was computed for each trajectory frame and not ensemble averaged we then calculate the ensemble average over the locations of the contour that were determined for each of the trajectory frames.

With the location of the contour on this spherical grid determined, we then use the fact that any finite valued function of the polar angles and its derivatives with respect to $\tht$ and $\ph$ can be expanded in spherical harmonics, $Y_{lm}\tp$:
\begin{equation}
\begin{aligned}
\dfrac{r \tp}{\mr} & = \sum_{l=0}^\infty \sum_{m=-l}^l R_{lm} Y_{lm}\tp \\ 
\frac{\partial (r/\mr)}{\partial \tht} & = \sum_{l=0}^\infty \sum_{m=-l}^l R_{lm} \frac{\partial Y_{lm}\tp}{\partial \tht} \\
\frac{\partial (r/\mr)}{\partial \ph} & = \sum_{l=0}^\infty \sum_{m=-l}^l R_{lm} i m Y_{lm}\tp 
\end{aligned}
\label{eqn:harmonic}
\end{equation}
The $R_{lm}$ values in these expressions are given by the following integral, which we can perform numerically using the finite set of values that we obtained for $ r \tp/\mr $:
\begin{equation}
R_{lm} = \int_{-\pi}^\pi \int_0^\pi \dfrac{r \tp}{\mr}  Y_{lm}^* \sin(\tht) \textrm{d}\ph\textrm{d}\tht.
\label{eqn:sp-coeff}
\end{equation}
In fact, this process is made even more straightforward as the symmetry of the lattice ensures that many of these $R_{lm}$ coefficients are zero.

The $R_{lm}$ values that we determine by performing these integrals can be inserted into the summations in Eqn.~\eqref{eqn:harmonic} and these summations can, in turn, be inserted into Eqn.~\eqref{eqn:final-g} to get an expression for $g\tp/\gamma$ in terms of spherical harmonics.  $g\tp/\gamma$, however, is just a finite valued function of the polar angles.  As such, it too can be expanded in spherical harmonics using a similar expression to~\eqref{eqn:harmonic}. 
As shown in the \si{}, when one substitutes the expansions for $g/\gamma$ and $r/\bar{r}$ in the linearized Eqn.~\eqref{eqn:final-g}, it is possible to equate the spherical harmonics coefficients in these two expansions and to obtain the following expression linking $G_{lm}$ and $R_{lm}$:
\begin{multline}
G_{lm} = R_{lm} +  \dfrac{1}{2}\sum_{l_1=0}^\infty \sum_{m_1=-l_1}^{l_1} \sum_{l_2=0}^\infty \sum_{m_2=-l_2}^{l_2} \\
R_{l_1 m_1} R_{l_2 m_2}\left( T^{lm}_{l_1m_1l_2m_2} + K^{lm}_{l_1m_1l_2m_2} \right)
\label{eqn:g-sph-harm}
\end{multline}
The constants
\begin{multline}
        T^{lm}_{l_1m_1l_2m_2}  = \\
        \int_{-\pi}^\pi \int_0^\pi \frac{\partial Y_{l_1m_1}\tp}{\partial \tht} \frac{\partial Y_{l_2 m_2}\tp}{\partial \tht} Y^*_{lm} \sin(\tht) \textrm{d}\ph \textrm{d}\tht
\end{multline}
and
\begin{multline}
K^{lm}_{l_1m_1l_2m_2}  = \\
-\int_{-\pi}^\pi \int_0^\pi \frac{m_1 m_2 Y_{l_1 m_1}\tp Y_{l_2 m_2}\tp }{\sin^2(\tht)} Y^*_{lm} \sin(\tht) \textrm{d}\ph \textrm{d}\tht ,
\end{multline}
in these expressions can be calculated analytically.

In the remainder of this paper, Eqn.~\eqref{eqn:g-sph-harm} is used to obtain \gt{coefficients for the spherical harmonics in the linear expansion for the} anisotropy in the surface tension for a crystalline nucleus of Lennard Jonesium.  This expression is then used to calculate the relative free energies of various surfaces of this crystal. The values obtained for these relative free energies are then compared with literature values for the anisotropy that were obtained by performing simulations of planar interfaces.    

\section{Simulation details}
\label{sec:details}

To test the methods discussed in Section~\ref{sec:method}, we performed MD simulations of homogeneous nucleation for a simple but realistic Lennard-Jones system. 
The simulations were identical to the ones in Ref.~\citenum{cheng2017gibbs}, so the NPT ensemble was employed throughout to simulate a supercell containing 23,328 atoms.  The Nose-Hoover thermostat was used to maintain the system at the melting temperature, $T_m=0.6185$, and an isotropic barostat was used to ensure that the pressure equalled zero. 
Twelve independent simulations of approximately $6 \times 10^6$ steps were performed with a time step of 0.004 Lennard-Jones time units and snapshots were stored every 5,000 MD steps.

In all these simulations, biased sampling using the well-tempered metadynamics protocol with adaptive Gaussians was performed.  Notice that other sampling techniques such as seeding~\cite{espinosa2016seeding,zimmermann2015nucleation,cheng2018theoretical} and umbrella sampling~\cite{torrie1977nonphysical,reinhardt2012free,cheng2018theoretical} could have been used instead of metadynamics.  What is more important is the collective variable (CV) the bias acts upon which in this case was the one employed in Ref.~\citenum{angioletti2010solid,cheng2015solid,cheng2017gibbs}; namely, $\ph = \sum_i S(\kappa_i)$.  In this expression, the rotational-variant order parameter $S(\kappa_i)$ that is calculated for atom $i$ considers the arrangement of the atoms in the first coordination sphere and determines whether these atoms are arranged as they would be in the solid.  
For atoms that are part of the solid nucleus in the centre of the box, this quantity thus approaches one.  For the atoms in the liquid that surrounds the nucleus, by contrast, the distribution of $S(\kappa_i)$ values is centred on zero.
It is important to note that the metadynamics bias only acts on the size of the crystal nucleus.  As it does not act upon the nucleus' shape, any ensemble averages related to the shape can thus be computed without reweighting~\cite{cheng2018communication}.  Furthermore, because the collective variable depends on the orientation of the crystal, the nucleus always has the same orientation in the simulation box. 

\begin{figure*}
  \includegraphics[width=\textwidth]{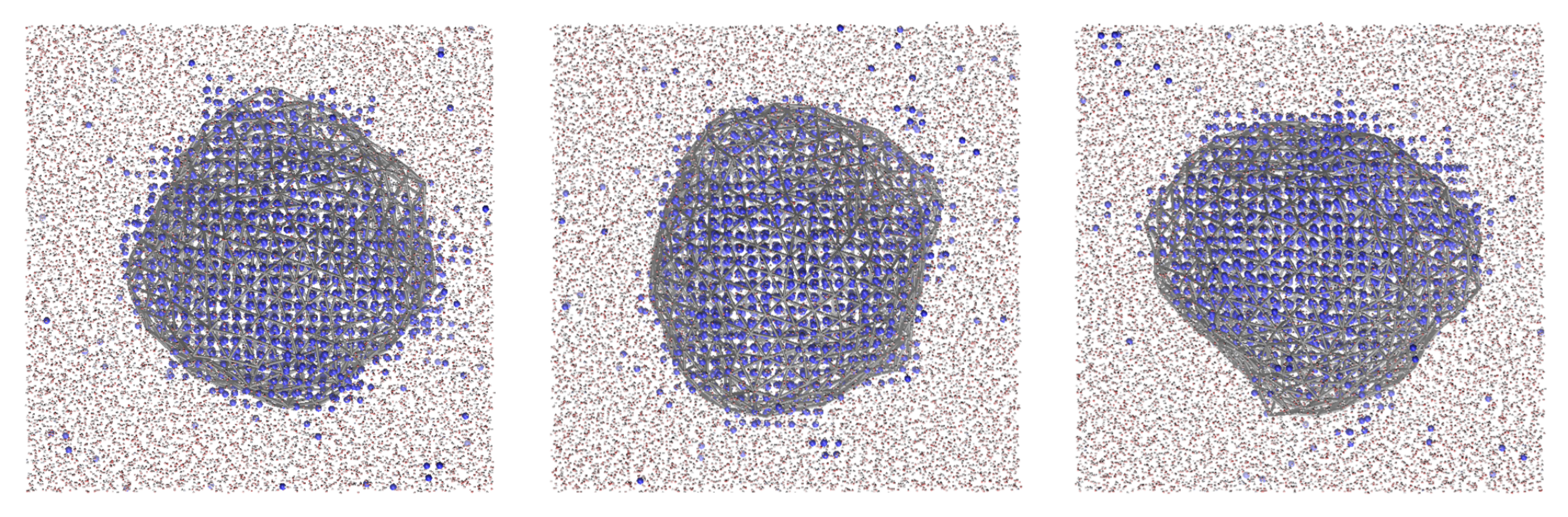}
      \caption{Snapshots of atomic configurations that contain a solid nucleus. In these figures, the atoms are coloured following the value of the atomic order parameter $S(\kappa_i)$ that was employed in Ref.~\citenum{cheng2015solid}. The grey wireframe indicates the instantaneous shape of the nucleus, which was determined by searching for the isocontour along the radial vectors on the Fibonacci sphere, as described in Section~\ref{sec:method}.}
          \label{fig:shape}
\end{figure*}

When analyzing the trajectories that were run at $T_m$, we picked out the 2035 trajectory frames that had a $\Phi=\sum_i\ph_i = \sum_i S(\kappa_i)$ value in a narrow window from 3760 to 3772. 
For $\Phi$ values of this magnitude, a Gibbs dividing surface based on $\ph$~\cite{cheng2016bridging} tells us that the crystal cluster contains about $3500$ atoms on average. 
Figure~\ref{fig:shape} shows the arrangement of atoms in a few of these trajectory frames. In these figures, the atoms are coloured following the value of the order parameter $S(\kappa_i)$.

For any choice of the atomic order parameter $\psi_i$, 
one can compute the phase-field $\tpsi\xyz$ using Eqn.~\eqref{eqn:pfield}.  In this work, the kernels in this expression were isotropic Gaussian kernels with a bandwidth of $\sigma_0=\sigma$, where $\sigma$ denotes the reduced Lennard-Jones unit of length.
The ensemble average of the phase-field $\avg{\tpsi\xyz}$ can be computed by averaging over the sampled trajectory frames because the value of the CV that was employed in the biased simulations depends on the orientation of the crystal. Consequently, all the frames are automatically aligned.  
We computed the ensemble average $\avg{\tpsi\xyz}$ on a $150\times150\times150$ grid of points.  \gt{Furthermore, suitable error bars on all averages were computed using block averaging.}  We found $ r \tp$ by finding the isosurface where the density was equal to 0.878~$\sigma^{-3}$ and evaluated the value of $r \tp$ at 377 grid points on a Fibonacci sphere.  
Much of this analysis was done using PLUMED~\cite{tribello2014plumed} and example input files have been uploaded to the PLUMED-nest~\cite{plumed-manifesto}.

\section{Analysis}
\label{sec:analysis}

\subsection{Instantaneous shape VS. average shape}

Before venturing into the spherical harmonics expansions and the Wulff constructions,
we first clarify the distinction between the instantaneous and the average shape of the interface.
In Section~\ref{sec:method}, we discussed two ways that an isosurface separating the solid and liquid phases can be determined from the trajectories.
If, for example, one takes the instantaneous values of $\{S(\kappa_i)\}$ for all the atoms in a single trajectory frame and constructs a phase-field $\tpsi\xyz$ using Eqn.~\eqref{eqn:pfield}, one can construct a reasonable representation for the instantaneous shape of the solid cluster by finding an isocontour in $\tpsi\xyz$ using Eqn.~\eqref{eqn:contour} and Eqn.~\eqref{eqn:gibbs} in conjunction. 
It is possible to find an instantaneous isocontour in this case because, for this particular choice of the atomic order parameter, the fluctuations in the value of $S(\kappa_i)$ for atoms in a solid/liquid are small compared to the difference between $\tpsis$ and $\tpsil$.
As a consequence, one can find a simply-connected isosurface that follows the shape of the nucleus
defined by the order parameter $S(\kappa_i)$ in the instantaneous phase field.

Figure~\ref{fig:shape} shows examples of these instantaneous isosurfaces for a few atomic configurations.  The instantaneous surface for the largest nucleus that has formed in these configurations is irregularly shaped and does not exhibit facets.
We note in passing that one could compute the amplitude of the fluctuations in the expansion of the surface in spherical harmonics, and use it to estimate the interfacial stiffness in a way that is analogous to the capillary fluctuation method~\cite{hoyt2001method}. Figure~\ref{fig:phi-comparison} shows that the amplitude of the fluctuations reflects the symmetry of $\gamma\tp$. We find, however, that the statistical errors in the fluctuations are larger than those in the average shape of the nucleus, and we thus did not attempt to verify the consistency of the fluctuation spectrum quantitatively. 

For these solid clusters, that contain about 3500 atoms, we find that the value of the atomic order parameters $S(\kappa_i)$ for the atoms in the centre of the cluster is similar to the value this quantity would take for an atom in bulk solid.  Furthermore, for those atoms that are close to the surface, the value of $S(\kappa_i)$ is between the value observed for the bulk solid and the bulk liquid.  In the surrounding liquid phase $S(\kappa_i)$ takes a value that is close to the value that it would take in the bulk liquid.  It is important to note that there are small clusters of atoms that the order parameter would indicate are solid-like in this part, but these same features would be seen in any bulk liquid phase.

If the atomic order parameter undergoes large fluctuations inside the bulk phases, or if the bandwidth of the kernel functions in Eqn.~\eqref{eqn:pfield} is set too small, the construction of a simply connected isosurface that separates the solid and liquid phases from the positions that the atoms take in a single trajectory frame becomes difficult and unstable. 
In these cases, one can only determine the shape of the nucleus by using the average value of the order parameter, i.e. 
by finding a contour that satisfies
$\avg{\tpsi\xyz} - \tpsio = 0$.  As discussed in more detail in appendix \ref{append:one}, we did not find any noticeable difference between the average shapes found using these two approaches when the order parameter $\{S(\kappa_i)\}$ was used to calculate the phase field. The fact that there is little difference is unsurprising though.  The number of solid atoms is prevented from changing by the restraint.  

\begin{figure}
    \centering
    \includegraphics[width=0.45\textwidth]{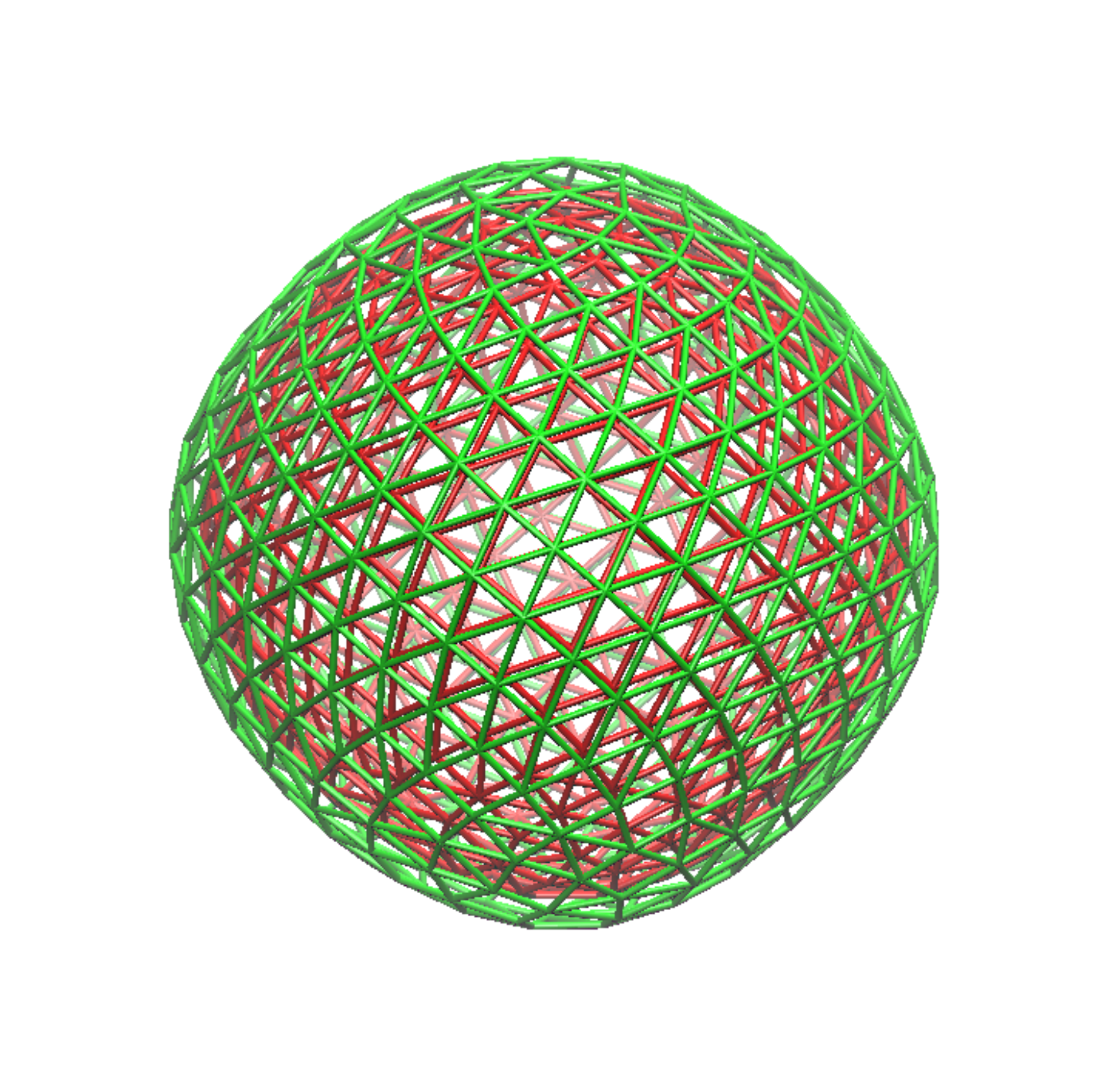}
    \caption{The average surface of the nucleus $r \tp$, which was determined using the methods describe in Sec.~\ref{sec:method}. The green wire frame shows the isosurface determined from the ensemble averaged phase field based on $\{S(\kappa_i)\}$. The red surface shows the equimolar isosurface that was calculated from the average atomic density field $\avg{\rho\xyz}$.}
    \label{fig:avgshape}
\end{figure}

Consequently, the position of the interface does not change by much from frame to frame.  The shape in each frame thus resembles the average shape $r \tp$ that is obtained from $\avg{\tpsi\xyz}$, and that is plotted using the green wireframe in Figure~\ref{fig:avgshape}.  As can see from this figure, the average shape is smooth and close to spherical.

\subsection{Dividing surfaces}

As discussed in Section~\ref{sec:method}, the number density of atoms should be used as the phase-field in Eqn.~\eqref{eqn:contour} as this ensures that the chemical potential difference per unit volume, $\mu$, is constant~\cite{tolman1948consideration,cheng2018communication}.
To construct this equimolar isosurface, we set each $\psi_i$ value equal to 1 when applying Eqn.~\eqref{eqn:pfield} to obtain $\rho\xyz$.  The change in density upon solidification is rather small for a system described by an LJ potential, and the local atomic density thus undergoes substantial fluctuations in both the liquid and solid phases. As a consequence, for this phase-field based on the density, and unlike the instantaneous phase-field based on $S(\kappa_i)$,
it is no longer possible to obtain a simply-connected instantaneous shape of the nucleus from a single trajectory frame.  An average phase field $\avg{\rho\xyz}$ must, therefore, be taken over multiple frames that all contain nuclei with similar sizes. 
An isocontour at $\rhoo=0.878 \sigma^{-3}$ in the average phase-field computed in this way is shown in red in figure \ref{fig:avgshape}.  This isocontour corresponds to the location of the equimolar interface because $\rhoo$ was set by following the zero surface excess condition in Eqn.~\eqref{eqn:gibbs}.  As can be seen in figure \ref{fig:avgshape}, and as is explored in more detail in appendix \ref{append:one}, the equimolar and the $S(\kappa_i)$-based surfaces differ.  The equimolar surface, in particular, is closer to the solid core.  As explained extensively in Ref.~\citenum{cheng2015solid}, these differences that arise because  of the Gibbs dividing surface that is being used can cause subtle changes in the value and anisotropy of $\gamma\tp$.

\subsection{Cubic Harmonics expansion}

\begin{figure*}
    \centering
    \includegraphics[width=0.95\textwidth]{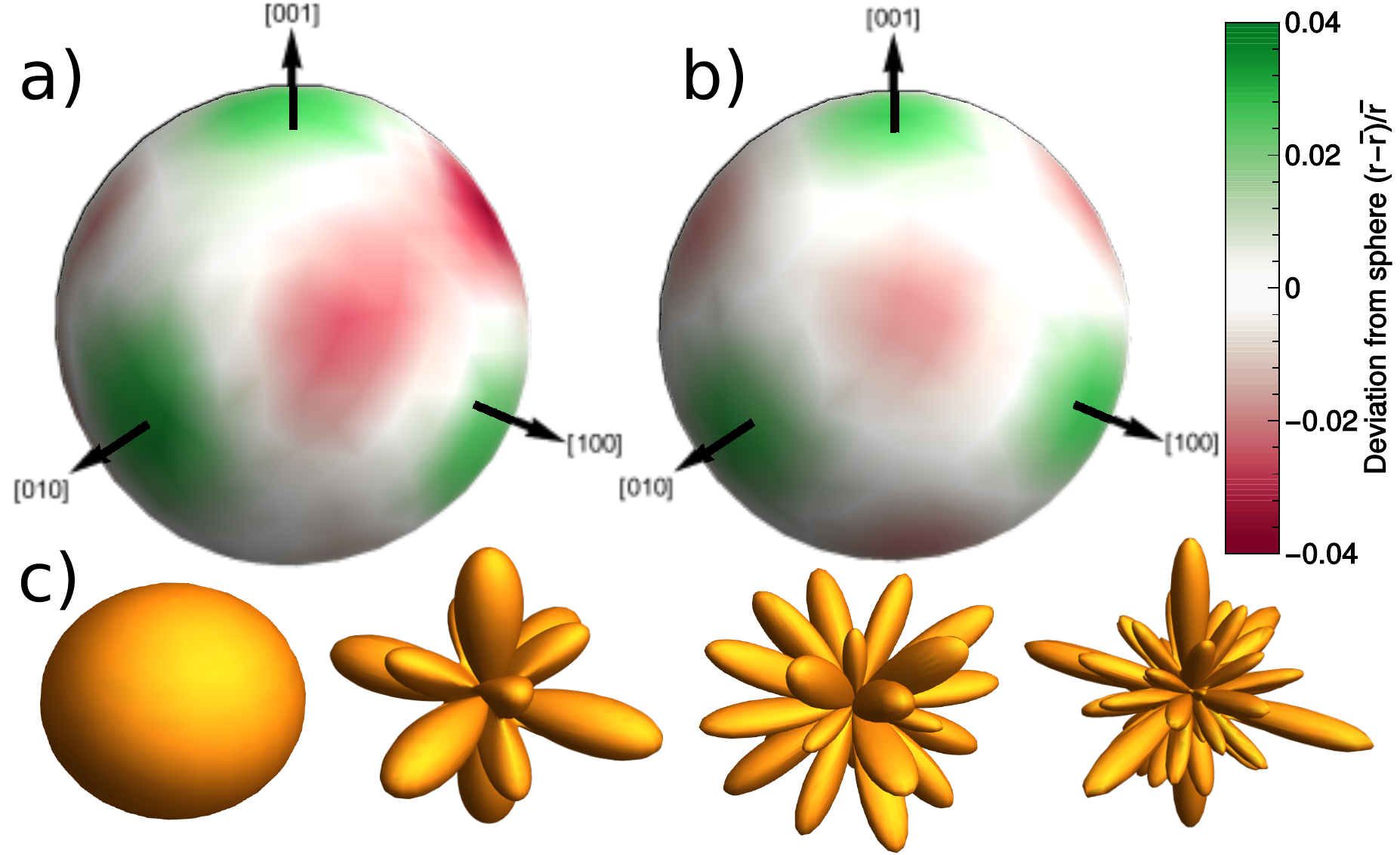}
    \caption{Illustrations of the average shape of the nucleus,
    and how it can be expanded in spherical harmonics.
    Panel a shows the $r \tp$ that was extracted from the atomic density field $\avg{\rho\xyz}$ that was computed from MD simulations using Eqn.~\eqref{eqn:contour}.
    Panel b shows an approximation to the average shape that is constructed using a sum of spherical harmonics with coefficients determined using Eqn.~\eqref{eqn:sp-coeff}.  As discussed in the text, the crystal symmetry ensures that the coefficients for many of the spherical harmonics are zero. We thus actually expand $r \tp$ in the cubic harmonics that have the appropriate symmetry.  Panel c, therefore, shows the particular set of cubic harmonic functions that were used. }
    \label{fig:ch}
\end{figure*}

Having obtained the equilibrium surface of the nucleus $r\tp$ from the average atomic density field $\avg{\rho\xyz}$,
we expanded the shape $r\tp/\mr$ in spherical harmonics to perform the reverse Wulff construction (Eqn.~\eqref{eqn:Wulff}) in a spherical harmonics basis (Eqn.~\eqref{eqn:g-sph-harm}).
The upper left panel of Figure~\ref{fig:ch} shows the shape $r\tp/\mr$ computed on the Fibonacci sphere grid.
The shape is approximately spherical, but it exhibits a small
anisotropy that is indicated more clearly by the colouring.
As shown in Eqn.~\eqref{eqn:harmonic} we can represent the shape $r \tp$  by using an expansion in the spherical harmonics with coefficients given by Eqn.~\eqref{eqn:sp-coeff}.  Furthermore, the fact that the face-centred-cubic (fcc) crystal structure of the Lennard-Jones solid has cubic symmetry ensures that the coefficients for many of these Spherical Harmonic functions are identically zero. 
In fact, incorporating this symmetry is simpler if one expands the shape in the Cubic Harmonics series~\cite{ALTMANN_1965,Muggli_1972,hoyt2001method,cheng2017gibbs}:
\begin{equation}
\begin{aligned}
X_{0}\tp = & Y_0^{0}\tp \\
X_{4}\tp = & Y_4^0\tp + \sqrt{\frac{5}{14}} \left[ Y_4^4\tp + Y_4^{-4}\tp \right] \\\
X_{6}\tp = & Y_6^0\tp - \sqrt{\frac{7}{2}} \left[ Y_6^4\tp + Y_6^{-4}\tp \right] \\
X_{8}\tp = & Y_8^0\tp + \sqrt{\frac{14}{99}} \left[ Y_8^4\tp + Y_8^{-4}\tp \right] \\
& + \sqrt{\frac{65}{198}} \left[ Y_8^8\tp + Y_8^{-8}\tp \right] \\
\end{aligned}
\label{eqn:cubic_harmonics}
\end{equation}
that are illustrated in the bottom panel of Figure~\ref{fig:ch}. These particular linear combinations  are real-valued functions that have the same symmetry as the fcc lattice.\gt{Together with higher-order terms, that we discard here, they form a complete basis to expand a function with a symmetry compatible with that of the lattice.}  %
Furthermore, because these functions are linear combinations of the spherical harmonics it is straightforward to extract the final coefficients that appear in Eqn.~\eqref{eqn:harmonic} from the coefficients of the basis functions in Eqn.~\eqref{eqn:cubic_harmonics}.

The upper right panel of Figure~\ref{fig:ch} illustrates the shape $r \tp/\mr$ that is obtained by performing the expansion of the isocontour shown in the figure's upper left panel using the spherical harmonics.  
Truncating the expansion and using a symmetry-adapted basis, ensures that this approximation for $r \tp/\mr$ is smooth and more exactly reflects the cubic symmetry of the system.  In other words, many artefacts that derive from the statistical noise resulting from the finite number of trajectory frames that were considered in the analysis are eliminated. 
It is worth noting that we estimated the random error due to limited sampling in our final estimates for the anisotropy in the interfacial free energy $g\tp/\gamma$, using the asymmetry in the computed $r \tp/\mr$.  
In other words, by truncating the expansion in this way, we are not disregarding the errors in our simulations altogether.  We are instead simply discarding components that we know to be zero given the symmetry of the system.

\begin{table}
    \centering
\begin{tabular}{c c c c }
\hline\hline
Coefficients & $X_4/X_0$   & $X_6/X_0$ & $X_8/X_0$   \\
\hline
$R_k$ &  0.00935 & -0.00034 & 0.00056 \\
$G_k$ & 0.00971 & -0.00031 & 0.00002 \\
\hline\hline
\end{tabular}
\vspace{3mm}
    \includegraphics[width=0.5\textwidth]{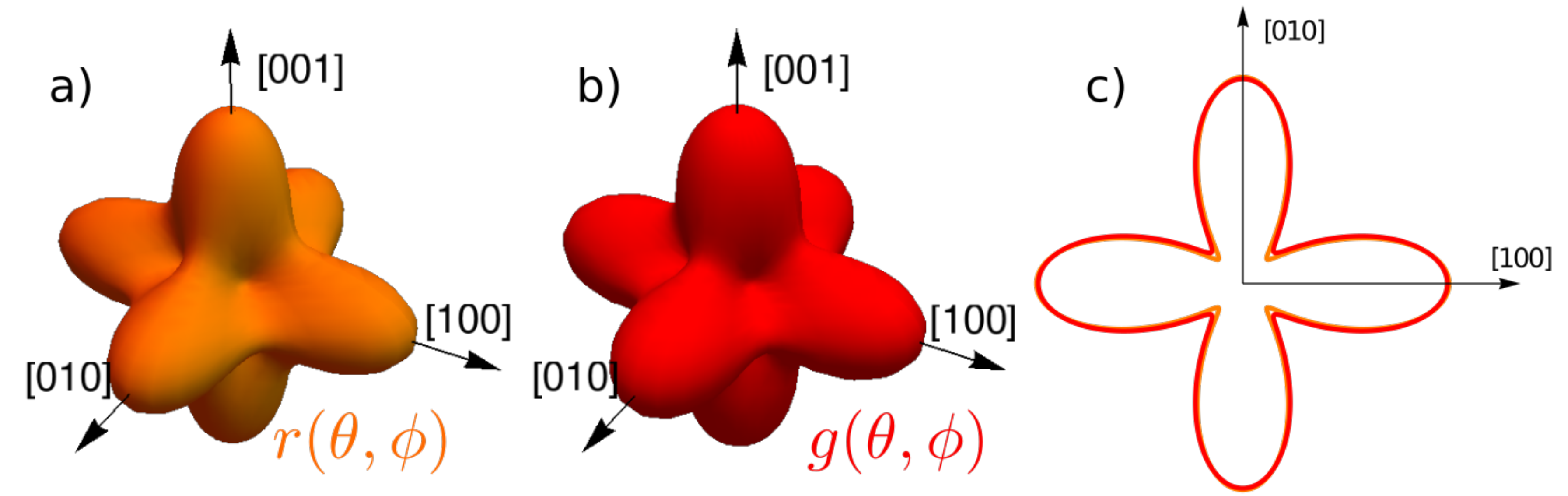}
    
\caption{
Information on the cubic harmonic expansions of $r\tp/\bar{r}$ ($R_k$) and $g\tp/\gamma$ ($G_k$).  The table gives the coefficients of the cubic harmonic expansion of $r\tp/\bar{r}$ ($R_k$) and $g\tp/\gamma$ ($G_k$), normalized by the coefficient of the isotropic component.  
The figures illustrate the anisotropies in the equilibrium shape of the nucleus and in the surface energy. 
    Panel a: anisotropy in $r \tp/\mr$. To make the anistropy more obvious, we plot $r \tp/\mr-0.96$.
    Panel b: anisotropy in $g\tp/\gamma$ Once again $g\tp/\gamma-0.96$ is shown here. 
    Panel c: A comparison between the anisotropies in $r \tp/\mr$ (orange curve) and $g\tp/\gamma$ (red curve) for the $(001)$ plane.
    This figure clearly shows most clearly the small differences between the anisotropies shown in a) and b). }
    \label{tab:aniso}
\end{table}

\section{Results\label{sec:results}}

As detailed in section \ref{sec:method}, we can extract an estimate for the anisotropy of the surface energy $g\tp/\gamma$ from the spherical harmonics expansion of $r \tp/\mr$ using Eqn.~\eqref{eqn:g-sph-harm}.  
The coefficients can then be converted (and symmetrized) into a cubic harmonic expansion, whose first coefficients are reported in Table~\ref{tab:aniso}.
The similarity in the coefficients is indicated by the similarity between the leftmost and middle panels of this figure.  The leftmost panel in this figure shows the anisotropy in the shape of the nucleus, $r/\mr$, while the middle panel shows the anisotropy in  $g\tp/\gamma$.
The small differences are highlighted in the right panel of the figure,
which shows cross-sections of these anisotropies between the surfaces in the $(001)$ plane at $z=0$. 
As was shown in Figure~\ref{fig:Wulfff}, these differences arise because of the (reverse) Wulff construction procedure. 
Ultimately, however, the fact that $r \tp/\mr$ and $g\tp/\gamma$ are so similar indicates that the nucleus is close to spherical.  

From the values of $g\tp/\gamma$ along  the $[100]$,$[111]$, and $[110]$ crystal lattice directions,
one can estimate the values of $g_{111}/g_{100}$ and $g_{110}/g_{100}$.  These values are reported in 
Table~\ref{tab:gmtm}
together with the the solid-liquid interfacial 
free energies that were computed along these three principle lattice directions in previous studies~\cite{cheng2018communication,cheng2015solid,becker2009atomistic} on this Lennard-Jones system.
Ref~\citenum{cheng2018communication,becker2009atomistic} used the capillary fluctuation method (CFM) to compute the surface tension,
while Ref.~\citenum{cheng2015solid,cheng2017gibbs} employed metadynamics simulations~\cite{laio-parr02pnas} to compute  the solid-liquid interfacial free energies along the three principal lattice directions.
In these previous studies, simulations of solid-liquid planar interfaces with the specified crystallographic orientations were performed.  In other words, these works did not investigate three-dimensional nuclei.
The planar interfaces were simulated in these studies~\cite{cheng2018communication,cheng2015solid,becker2009atomistic} by using a slab geometry so, once periodic boundary conditions are taken into account, the two-dimensional surface separating the solid from the liquid had infinite extent.
From these previous results, one can also compute the ratio between the surface energies for different directions that are reported in the fifth and sixth columns of the table.

The values of $g_{111}/g_{100}$ and $g_{110}/g_{100}$ in Table~\ref{tab:gmtm} computed using these different methods are all very similar.
Given the substantial difference between the geometry of the simulated system in this work and the geometries of the simulated systems in these other works, small discrepancies are to be expected and could be traced to finite-size effects, and minute differences in the computational setup.   
Nevertheless, the agreement between the free energies of planar interfaces and those computed from a statistical analysis of the 3D nucleus mean that the growth of a Lennard-Jones solid cluster within its melt is well described by classical nucleation theory.  Surface energies computed for planar interfaces can thus be used, together with an expansion of $\gamma\tp$ in a symmetry-adapted polar basis, to predict the equilibrium shape of the nucleus using the Wulff construction.

Even though instantaneous snapshots of the solid nucleus exhibit a rough surface, and relatively sharp edges, the average shape, which is the one that is most compatible with a thermodynamic description of the nucleation process, is  smooth and near-spherical, even though the solid cluster only contains a few thousand atoms.  Furthermore, the small anisotropies that are observed in the shape can be correctly predicted using the solid-liquid interfacial free energies calculated along high-symmetry lattice directions.
\gt{It would be interesting to perform a similar analysis on other systems, for example, cases where the faceting of the nucleus has been reported~\cite{bonati2018silicon,zhou2019observing},
or for smaller nuclei formed in deep undercoolings where the surface fluctuations were found to be important~\cite{prestipino2014shape}. Further analysis of these systems using the methods described in this article would allow one to determine whether this faceting is due to departures from classical nucleation theory and, more generally, whether inserting information on the anisotropy in the surface tension into CNT gives a complete description for the average shape of the nucleus.  }
In our view, taking suitable time averages or ensemble averages is an essential step in mapping the atomistic picture emerging from MD simulations into a phase-field or macroscopic picture.  This step is essential if one wishes  to comment on whether the results of an atomic-scale model are compatible with classical models. %

\begin{widetext}
\begin{table*}
\caption{A comparison of the computed interfacial free energy and anisotropy at $T_m$ from different studies.
The numbers in the brackets indicate the statistical uncertainties of the last digit.}
\label{tab:gmtm}
    \begin{tabular}{ c | c c c | c c }
   \hline\hline
  & \multicolumn{3}{ c| }{interfacial free energy} & \multicolumn{2}{ c }{anisotropy} \\ 
  Methods 
  & $g_{100}$ & $g_{111}$ & $g_{110}$
  & $g_{111}/g_{100}$ & $g_{110}/g_{100}$ \\
  
  \\\hline   
  This work & & & &
  0.951(3) & 0.967(2)
  \\
  Planar interface + CFM ~\cite{cheng2018communication} &
  0.365(2) & 0.350(2) & 0.355(2)
  & 0.959(5) & 0.973(5)
  \\
  Planar interface + Metadynamics ~\cite{cheng2015solid,cheng2017gibbs} &
  0.373(2) & 0.351(1) & 0.358(2)
  & 0.942(5) & 0.961(5)
  \\
  Planar interface + CFM ~\cite{becker2009atomistic} & 
  0.363(8) &  0.350(8) & 0.354(8)
  & 0.964 & 0.975
 \\ \hline \hline
    \end{tabular}
\end{table*}
\end{widetext}

\section{Conclusions}

We have shown how an estimate of the anisotropy in the surface tension for a crystalline nucleus can be extracted from a biased molecular dynamics trajectory.  The method that we have demonstrated works by using kernel density estimation to construct a phase-field representation.  This representation tells us whether or not the structure at each point in the simulation box is solid or liquid. We can thus locate the interface between these two phases by finding an isocontour in this phase field.  This isocontour can be used to construct an estimate of the average shape, $ r \tp$, of the nucleus.   The anisotropy in the surface tension can then be computed from this average shape by using ideas from classical nucleation theory.  
The framework presented in this paper of using isocontours of phase fields, with minor adaptions and extensions, 
could be applied to other scenarios including heterogeneous nucleation~\cite{pedevilla2018heterogeneous}, 
computation of contact angles~\cite{taherian2013contact},
and probing the structural heterogeneities in complex fluids~\cite{ansari2018high}.

When we do these surface excess free energy calculations for a system of Leonard Jones particles, we find that, even though instantaneous snapshots exhibit a rough, irregular surface, the average shape of the small nuclei that are formed in molecular dynamics simulations are smooth and symmetric.  %
We estimate the anisotropic surface energy $\gamma\tp$ by a reverse Wulff construction and obtain ratios for the surface tensions of high-symmetry crystal facets that are very similar to the ratios that are obtained when the surface tensions for each of these surfaces is calculated separately in a planar geometry.  We, therefore, argue that a Wulff construction that is parameterized using estimates of the surface energies that are computed from simulations of infinite period slabs provides a reasonable description of the average shape of a tiny nucleus of Lennard Jones. 
In other words, even though the crystal faces in these nuclei have only a small extent, they still share many of the properties of the infinite periodic surface. \gt{The morphology and stability of the crystalline nuclei composed of a few thousand atoms that we have observed in our simulations can thus be correctly described using macroscopic classical nucleation theory.} 

\appendix
\section{Representations in spherical coordinates}

The surface of a three-dimensional nucleus can be described using
\begin{equation}
     \xi (r, \tht, \ph) = r - r \tp =0,
\end{equation}
where $r$, $\tht$ and $\ph$ denote the radial distance, the polar angle and the azimuthal angle in the spherical coordinate system.
In other words, the function $r \tp$ fully characterizes star-shaped surfaces.

At each point $(r, \tht, \ph)$, the normalized normal vector $\vec{n}$ to the surface is
\begin{multline}
       \vec{n}=\dfrac{\nabla \xi}{| \nabla \xi |}
     =\dfrac{
     \pd{\xi}{r}\vec{e}_r-\dfrac{1}{r}\pd{\xi}{\tht}\vec{e}_{\tht}-\dfrac{1}{\sin(\tht)r}\pd{\xi}{\ph}\vec{e}_{\ph}
     }{
     \sqrt{(\pd{\xi}{r})^2+(\dfrac{1}{r}\pd{\xi}{\tht})^2+(\dfrac{1}{\sin(\tht)r}\pd{\xi}{\ph})^2}}\\
     =\dfrac{
     \vec{e}_r-\dfrac{r_{\tht}}{r}\vec{e}_{\tht}-\dfrac{r_{\ph}}{\sin(\tht)r}\vec{e}_{\ph}
     }{
     \sqrt{1+
     \dfrac{r^2_{\tht}}{r^2}+
     \dfrac{r^2_{\ph}}{\sin^2(\tht)r^2}}},  
\end{multline}
where $\vec{e}_{r}$, $\vec{e}_{\tht}$ and $\vec{e}_{\ph}$ are the unit vectors in the right-handed spherical coordinate system, and where
$r_{\tht}=\drtheta$ and $r_{\ph}=\drphi$ are partial derivatives of the function $r\tp$. 
We can express $\vec{n}$ using spherical coordinates $n=(1, \htht,\hph)$, where
\begin{multline}
         \htht=\tan^{-1} \left(
     \dfrac
    {\sqrt{(\sin(\tht)-\cos(\tht)\dfrac{r_{\tht}}{r})^2+\dfrac{r^2_{\ph}}{\sin^2(\tht)r^2}}}
     {\cos(\tht)+\sin(\tht)\dfrac{r_{\tht}}{r}}
     \right),\\
          \hph=\tan^{-1} \left(
     \dfrac
          {\sin(\ph)-\dfrac{\sin(\ph)\cos(\tht)}{\sin(\tht)}\dfrac{r_{\tht}}{r}-\dfrac{\cos(\ph)}{\sin^2(\tht)}\dfrac{r_{\ph}}{r}}
     {\cos(\ph)-\dfrac{\cos(\ph)\cos(\tht)}{\sin(\tht)}\dfrac{r_{\tht}}{r}+\dfrac{\sin(\ph)}{\sin^2(\tht)}\dfrac{r_{\ph}}{r}}
     \right).
     \label{eqn:hat}
\end{multline}

In the spherical coordinate system, it is also easy to express the surface area for each differential element $d \Omega = \sin(\tht) d \tht d \ph$ as:
\begin{equation}
    A(d \Omega) = \dfrac{|\vec{e}_r|}{\vec{n} \cdotp \vec{e}_r} r^2 d \Omega
    =r^2\sqrt{1+\dfrac{r^2_{\tht}}{r^2}+\dfrac{r^2_{\ph}}{\sin^2(\tht)r^2}}
     d \Omega,
     \label{eqn:a-surfacearea}
\end{equation}
The corresponding surface energy per unit area for this element is $g(\vec{n})$, where the vector $\vec{n}$ points in the direction $\htp$ as indicated in Figure~\ref{fig:Wulfff}.
Multiplying the surface area in Eqn.~\eqref{eqn:a-surfacearea} with the specific surface energy $g\htp$ and integrating over the sphere $\int_{\Omega} d\Omega$,
gives the surface energy of the nucleus that appears in Eqn.~\eqref{eqn:f}.

\section{Conditions for the equilibrium surface $r\tp$}

\begin{figure*}
    \centering
    \includegraphics[width=\textwidth]{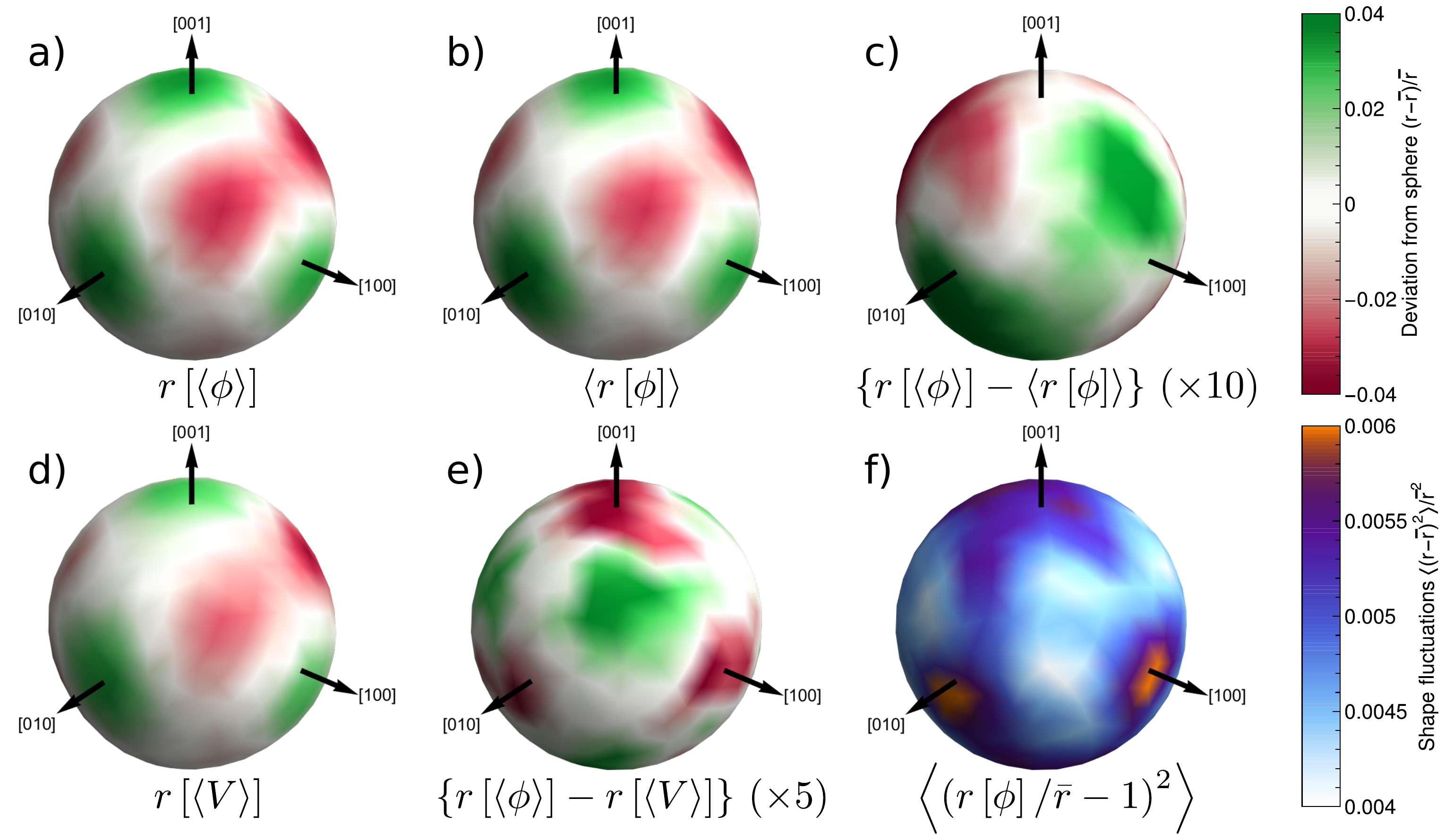}
    \caption{A comparison between the various methods that can be used to compute the location of the dividing surface.  The top row shows images of the dividing surface that were constructed using the order parameter.  Panel (a) shows an isocontour in a field that measures the average value of the order parameter.  Panel (b), by contrast, was constructed by finding the instantaneous location of the contour in each of the trajectory frames and by then computing an ensemble average of the contour location.  Finally, panel (c) shows that the differences between these two contours is tiny and is largely random. In the second row panel (d) shows the shape that is found by searching for an isocontour in the average density field.  Panel (e) then shows that there are substantial (non-random) differences between this shape and the shape that is found by searching for an isocontour in the field that measures the average value of the order parameter.  Lastly, panel (f) illustrates the magnitude of the fluctuations in the instantaneous shape at each point on the surface.}
    \label{fig:phi-comparison}
\end{figure*}

To make the derivations in this section easier to follow, we introduce the following shorthands:  
\begin{equation}
    t=\dfrac{r_{\tht}}{r}, \quad
    p=\dfrac{r_{\ph}}{\sin(\tht)r}, \quad
    g=g\htp.
\end{equation}
When a nucleus with a fixed size has its equilibrium surface $r\tp$, 
the first order derivatives of the free energy in Eqn.~\eqref{eqn:f} vanish:
\begin{equation}
   \pd{f}{r}=0, \quad 
   \pd{f}{p}=0, \quad
   \pd{f}{t}=0.
   \label{eqn:minf}
\end{equation} 
In the remainder of this appendix we will prove that these conditions lead to Eqn.~\eqref{eqn:Wulff},
which is equivalent to the expression for the Wulff construction.

By taking the derivative of Eqn.~\eqref{eqn:minf} we arrive at:
\begin{equation}
    0= \int_{\Omega} d\Omega  r\left(
    \dfrac{t}{\sqrt{1+c}}g
    +\sqrt{1+c}\pd{g}{t}
    \right),
\end{equation}

\begin{equation}
    0= \int_{\Omega} d\Omega  r\left(
    \dfrac{p}{\sqrt{1+c}}g
    +\sqrt{1+c}\pd{g}{p}
    \right),
\end{equation}

and

\begin{equation}
     0= \int_{\Omega} d\Omega \left( -\mu r^2
     +r\dfrac{2+c}{\sqrt{1+c}}g
     -r\sqrt{1+c}(t\pd{g}{t}+p\pd{g}{p})
     \right).
\end{equation}
By eliminating $\pd{g}{t}$ and $\pd{g}{p}$ in the third equation using the first two, one can obtain the expression for $g\htp$ in Eqn.~\eqref{eqn:Wulff}.

\section{Manipulating $g\htp$}

This appendix explains how Eqn.~\eqref{eqn:final-g} can be derived from Eqn.~\eqref{eqn:inter-g}.
We take a Taylor expansion of $(\htht,\hph)$ in Eqn.~\eqref{eqn:hat} to the second order in $(t,p)$,
and obtain
\begin{equation}
\begin{aligned}
    \htht &=\tht - t
     +\dfrac{\cos(\tht)}{2\sin(\tht)} p^2 
    +\bigo{tp^2},\\
    \hph &=\ph - \dfrac{p}{\sin(\tht)}
     -\dfrac{\cos(\tht)}{\sin^2(\tht)} pt
    +\bigo{t^2p}.
\end{aligned}
\end{equation}
We drop second and higher order terms in $t,p$ and thereafter use
\begin{equation}
    \htht\approx \tht-\dfrac{1}{r} \pd{r}{\tht}, \quad
    \hph \approx \ph-\dfrac{1}{r\sin^2(\tht)}\pd{r}{\ph}.
\end{equation}
These relations ensure that one can express $r\tp$ by performing a first-order Taylor series around $\htp$:
\begin{equation}
    r\tp = r\htp + \dfrac{1}{r} \pd{r}{\tht}\times\pd{r}{\tht}
    +\dfrac{1}{r\sin^2(\tht)}\pd{r}{\ph} \times \pd{r}{\ph}
    +\bigo{\pds{r}{\tht}}.
    \label{eqn:expand-r}
\end{equation}
In addition, $\pd{r}{\tht}$ and $\pd{r}{\ph}$ can also be expanded in a similar fashion.
Combining these expansions with Eqn.~\eqref{eqn:inter-g} and dropping higher-order terms gives
\begin{equation}
    \dfrac{g\htp}{\gamma} = 
\frac{1}{\mr} 
\left\{ r\htp + \frac{1}{2 r\htp}\left[ \left(\pd{r}{\htht} \right)^2 +  \frac{1}{\sin^2(\htht) } \left(\frac{\partial  r}{\partial \hph} \right)^2 \right] \right\},
\label{eqn:expand-2}
\end{equation}
To obtain equation \ref{eqn:final-g} one must perform a change of variable from $\htp$ to $\tp$.  Then, finally, because the average shape of the nucleus is very close to a sphere, i.e.
$\delta r \equiv | r - \mr | \ll \mr$, 
one Taylor expands the expression in Eqn.~\eqref{eqn:expand-2} around $\mr$. 

\section{Spherical harmonic expansion of equilibrium shape and anisotropy}

\gt{
The first step in deriving Eqn.~\eqref{eqn:g-sph-harm} is to perform a spherical harmonics expansion of both sides of Eqn.~\eqref{eqn:final-g}.  When this procedure is complete the left-hand side is just
\begin{equation}
    G_{lm}=\int_{\tht=0}^{\pi}\int_{\ph=0}^{2\pi} \dfrac{g\tp}{\gamma} Y^{\star}_{lm}\sin(\tht) d \tht d\ph,
\end{equation}
The first term on the right hand side, meanwhile, is $R_{lm}$.  By using the results in Eqn.~\eqref{eqn:harmonic} it is then possible to show that the second term is:
 \begin{multline}
       \dfrac{1}{2}\int_{\tht=0}^{\pi}\int_{\ph=0}^{2\pi}       \left[\sum_{l=0}^\infty \sum_{m=-l}^l R_{lm} \frac{\partial Y_{lm}\tp}{\partial \tht}\right]^2 Y^{\star}_{lm}\sin(\tht) d \tht d\ph
 \\=
     \dfrac{1}{2}\sum_{l_1=0}^\infty \sum_{m_1=-l_1}^{l_1} \sum_{l_2=0}^\infty \sum_{m_2=-l_2}^{l_2} 
R_{l_1 m_1} R_{l_2 m_2}\\
\times \int_{-\pi}^\pi \int_0^\pi \frac{\partial Y_{l_1m_1}\tp}{\partial \tht} \frac{\partial Y_{l_2 m_2}\tp}{\partial \tht} Y^*_{lm} \sin(\tht) \textrm{d}\ph \textrm{d}\tht
 \end{multline}
 A similar procedure can be used to expand the third term 
and hence to finally obtain Eqn.~\eqref{eqn:g-sph-harm}.
}

\section{Analysis of the various ways of constructing the dividing surface}
\label{append:one}

As discussed in the main text, we computed the shape of the nucleus in several different ways.  In particular, we used two different order parameters for the extensive quantity $\Psi$ that is introduced when we discuss equation \ref{eqn:pfield}; namely, the order parameter $\Phi$ and the volume.  For the analysis that was performed using $\Phi$ we computed the ensemble average of the phase-field before searching for the contour and also found the instantaneous contour for each of our trajectory frames and then computed the final, average shape by computing an ensemble average over these instantaneous ones.  
Figure \ref{fig:phi-comparison} shows a comparison of the shapes that were obtained using these various methods.  Panel (c) illustrates that there is little difference between the shape obtained by searching for the isocontour in the field that measures the average value of the order parameter and the average for all the instantaneous contours.  Furthermore, the differences there are between these two shapes appear to be due to statistical noise and are thus unlikely to affect the final result.  It is thus acceptable to use the computationally cheaper and more stable approach of only searching for the final contour in the average order parameter field computed from all the frames sampled.  

The fact that one can use this approach of searching for the contour in an averaged field is useful because, as discussed in the main text, the large fluctuations in the average volume per atom make it impossible to calculate instantaneous isosurfaces in the instantaneous density field.  The equimolar shape that is shown in figure \ref{fig:phi-comparison}(d) was thus computed from a density field that was computed by taking an average over all the sampled frames.  As you can see from figure \ref{fig:phi-comparison}(e) there are substantial (non-random) differences between this equimolar shape and the shape that was found by searching for an isocontour in the field that measures the average value of the order parameter.  As discussed in the main text, these differences are essential as many assumptions within classical nucleation theory are predicated on the assumption that the excess volume of the interface is zero, which is only true of the equimolar dividing surface.

Figure \ref{fig:phi-comparison}(f) shows the magnitude of the fluctuations in the shape of the nucleus for the trajectory.  This figure was constructed by analysing the instantaneous isocontours in the fields that measured the instantaneous value of the order parameter at each point in the simulation cell. Our simulations were not run for long enough to converge these fluctuations.  If one were to converge the fluctuations in the shape, one could compute the surface stiffness and hence the surface tension using ideas from capillary wave theory.

\section{Using Wulff construction to predict the average shape of the nucleus from planar interfacial free energies}
\gt{
In the main manuscript we showed that the reverse Wulff construction (Eqn.~\eqref{eqn:g-sph-harm}) can be derived from the linearized form of Eqn.~\eqref{eqn:final-g}.  Furthermore, we showed that this reverse Wulff construction allowed one to compute the surface tensions from the average shape.  One can also use Eqn.~\eqref{eqn:final-g}, however, to derive the 
Wulff construction.  In other words, one can use this equation to derive the following expression, which relates the coefficients, $R_{lm}$, of the spherical harmonics in the linear expansion for the shape, $r/\bar{r}$, to the coefficients of the spherical harmonics in the linear expansion for the surface tension, $g/\gamma$:
\begin{multline}
R_{lm} = G_{lm} -  \dfrac{1}{2}\sum_{l_1=0}^\infty \sum_{m_1=-l_1}^{l_1} \sum_{l_2=0}^\infty \sum_{m_2=-l_2}^{l_2} \\
R_{l_1 m_1} R_{l_2 m_2}\left( T^{lm}_{l_1m_1l_2m_2} + K^{lm}_{l_1m_1l_2m_2} \right),
\label{eqn:wulff-sph}
\end{multline}

This expression is useful if one already has values for $g/\gamma$, from, for instance, planar interface simulations~\cite{cheng2015solid,cheng2017gibbs}. When the values of these quantities are inserted into Eqn.~\eqref{eqn:wulff-sph}, an approximate expression for the   average shape of the nucleus can be determined.  To further demonstrate that classical nucleation theory is valid for this particular system we thus took the three values of $g_{100}$,$g_{111}$ and $g_{110}$  from Ref.~\citenum{cheng2015solid,cheng2017gibbs} (Table~\ref{tab:gmtm}) that were obtained by performing simulations of planar interfaces.  From these surface tensions, we obtained the following ratios for the cubic harmonics coefficients: $X_4/X_0=0.01175$ and $X_6/X_0=-0.00045$.
 Eqn.~\eqref{eqn:wulff-sph} and a spherical harmonics expansion was then used to calculate the average shape $r_p/\mr$. The final result we obtained is shown in Fig.~\ref{fig:compare-shape}b.  In that figure, we also show the average shape $r \tp/\mr$ that we computed from the nucleation simulations in this paper.  As you can see the two shapes shown in Fig.~\ref{fig:compare-shape} are very similar, which is further evidence that it is appropriate to describe the nucleation of this system using CNT.

\begin{figure}
    \centering
    \includegraphics[width=\columnwidth]{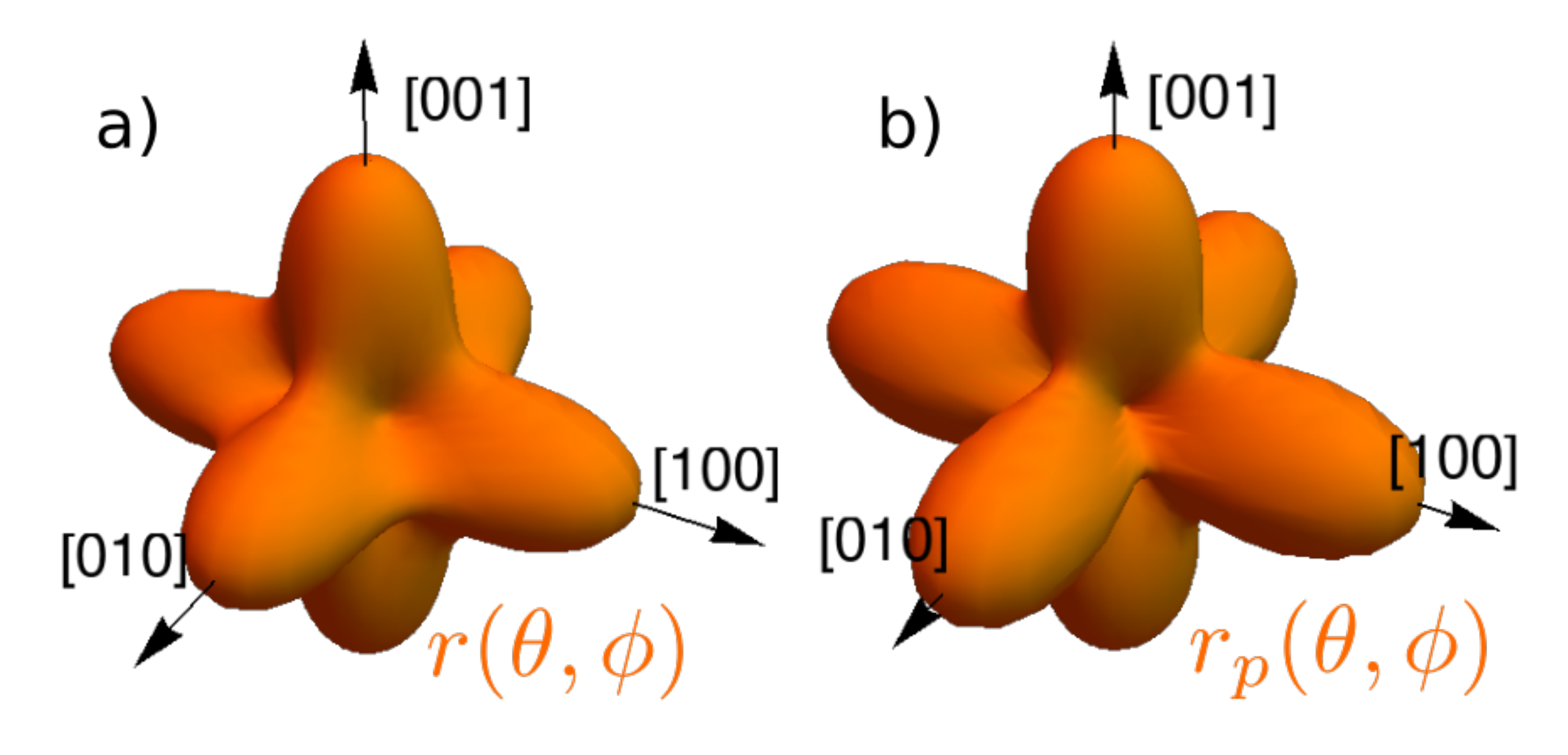}
    \caption{A comparison of the shape of the nucleus, $r \tp/\mr$, that is predicted by using a Wulff construction that is parameterized using the surface tensions for the $g_{100}$,$g_{111}$ and $g_{110}$ that were computed from the simulations of planar interfaces conducted in Ref.~\citenum{cheng2015solid,cheng2017gibbs} (panel b) and the average shape that was observed in our simulations (panel a). 
    In both panels we plot $r \tp/\mr-0.96$ so as to make it easier to see the anisotropy.}
    \label{fig:compare-shape}
\end{figure}
}
\begin{acknowledgments}
BC acknowledges funding from Swiss National Science Foundation (Project P2ELP2-184408).
BC acknowledges the resources provided by the Cambridge Tier-2 system operated by the University of Cambridge Research Computing Service (http://www.hpc.cam.ac.uk) funded by EPSRC Tier-2 capital grant EP/P020259/1.
\end{acknowledgments}

\end{document}